# Correlated noise in a logistic growth model

Bao-Quan Ai, Xian-Ju Wang, Guo-Tao Liu, and Liang-Gang Liu*
*Department of Physics, ZhongShan University, GuangZhou, People's Republic of China*

The logistic differential equation is used to analyze cancer cell population, in the presence of a correlated Gaussian white noise. We study the steady state properties of tumor cell growth and discuss the effects of the correlated noise. It is found that the degree of correlation of the noise can cause tumor cell extinction.

## I. INTRODUCTION

Recently, nonlinear stochastic systems with noise terms have attracted extensive investigations and the concept of noise-induced transition has got wide applications in the field of physics, chemistry, and biology [1,2]. Usually, in these systems the noise affects the dynamics through a system variable, i.e., the noise is both multiplicative and additive [3]. The focal theme of these investigations is to study the steady state properties of systems in which fluctuations, generally applied from outside, are considered independent of the system's characteristic dissipation. Since the two types of fluctuations have a common origin, they are correlated in the relevant time scale of the problem [4]. On the level of a Langevin-type description of a dynamical system, the presence of correlation between noises can change the dynamics of the system [5,6]. Correlated noise processes have found applications in a broad range of studies such as steady state properties of a single mode laser [7], bistable kinetics [8], directed motion in spatially symmetric periodic potentials [9], stochastic resonance in linear systems [10], and steady state entropy production [11]. In this paper we study a tumor cell growth model in the presence of correlated additive and multiplicative noise and show how noise correlation can dynamically cause tumor cell extinction.

## II. THE TUMOR CELL GROWTH MODEL

The logistic growth model has been used in many cases as a basic model of both cell growth and, more particularly, tumor cell growth [12,13]. Here, we only consider tumor cell growth. The logistic differential equation is shown,

$$\frac{dx}{dt} = ax - bx^2, \qquad (1)$$

where $x$ is the tumor mass, $a$ the growth rate, and $b$ the cell decay rate. We consider effects due to some external factors such as temperature, drugs, radiotherapy, etc. These factors can influence the tumor mass directly as well as alter the tumor growth rate. In other words, the fluctuation of these factors affects the parameter $a$ generating multiplicative noise and, at the same time, some factors, such as drugs and radiotherapy, restrain the number of tumor cells, giving rise to a negative additive noise. As a result, we obtain

$$\frac{dx}{dt} = ax - bx^2 + x\epsilon(t) - \Gamma(t), \qquad (2)$$

where $\epsilon(t)$ and $\Gamma(t)$ are Gaussian white noises with the following properties:

$$\langle \epsilon(t) \rangle = \langle \Gamma(t) \rangle = 0, \qquad (3)$$

$$\langle \epsilon(t)\epsilon(t') \rangle = 2D\delta(t-t'), \qquad (4)$$

$$\langle \Gamma(t)\Gamma(t') \rangle = 2\alpha\delta(t-t'), \qquad (5)$$

$$\langle \epsilon(t)\Gamma(t') \rangle = 2\lambda\sqrt{D\alpha}\delta(t-t'), \qquad (6)$$

where $\alpha$ and $D$ are the strengths of the two noises and $\lambda$ denotes the degree of correlation between $\epsilon(t)$ and $\Gamma(t)$ with $0 \leq \lambda < 1$.

## III. STEADY STATE ANALYSIS AND RESULTS

Since the cell number $(x)$ cannot be negative, we can derive the Fokker-Planck equation for the evolution of steady probability distribution function (SPDF) corresponding to Eq. (2) under the constraint $x \geq 0$. The equation is [14]

$$\frac{\partial P(x,t)}{\partial t} = -\frac{\partial A(x)P(x,t)}{\partial x} + \frac{\partial^2 B(x)P(x,t)}{\partial x^2}, \qquad (7)$$

where $P(x,t)$ is the probability density and

$$A(x) = ax - bx^2 + Dx - \lambda\sqrt{D\alpha}, \qquad (8)$$

$$B(x) = Dx^2 - 2\lambda\sqrt{D\alpha}x + \alpha. \qquad (9)$$

The stationary probability distribution of equation is given as [14]

$$P_{st}(x) = \frac{N}{B(x)} \exp\left[\int^x \frac{A(x')dx'}{B(x')}\right], \qquad (10)$$

where $N$ is a normalization constant. Using the explicit forms of $A(x)$ and $B(x)$ we obtain the following SPDF [15]:

---
*Email address: stdp05@zsu.edu.cn

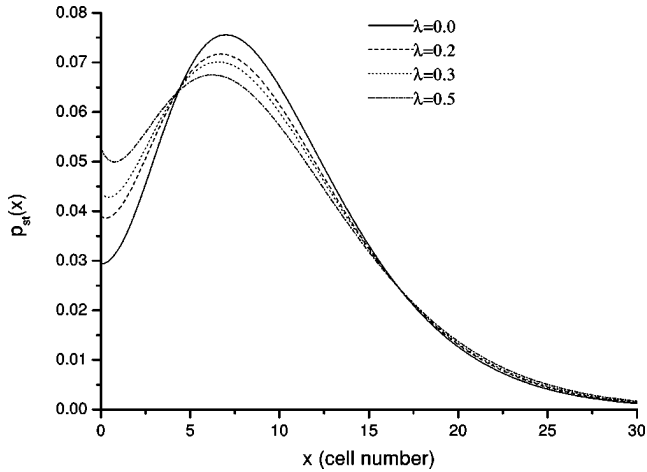

FIG. 1. Plot of $P_{st}(x)$ (probability density) vs $x$ (cell number) for low values of the noise-noise correlation $\lambda$. $D=0.3$, $\alpha=3.0$, $a=1$, $b=0.1$ and $\lambda=0$, 0.2, 0.3, and 0.5, respectively (units are arbitrary).

$$P_{st}(x) = NB(x)^{C-1/2} \exp\left\{ f(x) + \frac{E}{\sqrt{D\alpha(1-\lambda^2)}} \right.$$
$$\left. \times \arctan\left( \frac{Dx - \lambda\sqrt{\alpha D}}{\sqrt{D\alpha(1-\lambda^2)}} \right) \right\} \quad (0 \leq \lambda < 1), \quad (11)$$

where

$$C = \frac{a - 2\lambda\sqrt{\frac{\alpha}{D}}b}{2D}, \quad f(x) = -\frac{b}{D}x, \quad (12)$$

$$E = b\frac{\alpha}{D} - \left(a + 2\lambda\sqrt{\frac{\alpha}{D}}b\right)\lambda\sqrt{\frac{\alpha}{D}}. \quad (13)$$

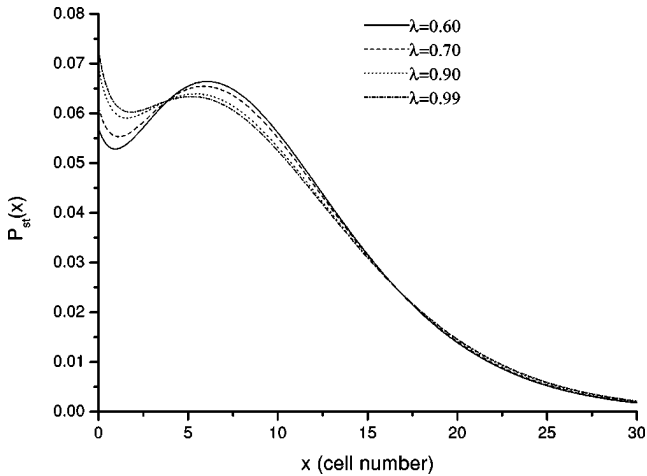

FIG. 2. Plot of $P_{st}(x)$ (probability density) vs $x$ (cell number) for high values of the noise-noise correlation $\lambda$. $D=0.3$, $\alpha=3.0$, $a=1$, $b=0.1$ and $\lambda=0.60$, 0.70, 0.90, and 0.99, respectively (units are arbitrary).

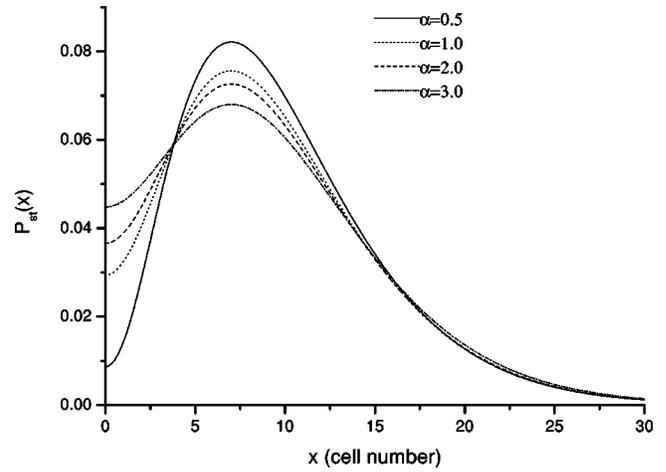

FIG. 3. Plot of $P_{st}(x)$ (probability density) vs $x$ (cell number) for different values of the additive noise intensity $\alpha$. $a=1$, $b=0.1$, $D=0.3$, $\lambda=0.0$ and $\alpha=0.5$, 1.0, 2.0, and 3.0, respectively (units are arbitrary).

The extrema of $P_{st}(x)$ obey a general equation $A(x) - [dB(x)/dx] = 0$:

$$bx^2 + (D-a)x - \lambda\sqrt{D\alpha} = 0. \quad (14)$$

If $\lambda = 0$ the last term of the Eq. (14) vanishes and we have the extrema of SPDF for only multiplicative noise processes. In fact, for zero correlation, the additive noise has no effect on the position of the extrema of SPDF which are $x=0$ and $x=(a-D)/b$.

In Figs. 1 and 2, we show the effect of the correlation parameter $\lambda$ on the steady state probability distribution (SPD). As the value of $\lambda$ increases, $P_{st}(x)$ increases at small $x$, and decreases at large $x$. Since $x$ denotes the tumor cell population, it is clear that increasing $\lambda$ tumor cell population disappears. In other words, the distribution of cell population which was mainly peaked about zero (for a large value of $\lambda$) signifying high extinction rates, moves toward zero with the increase of the correlation parameter $\lambda$.

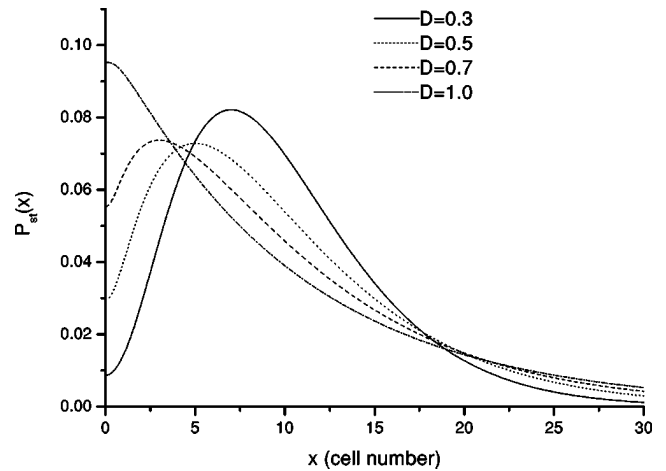

FIG. 4. Plot of $P_{st}(x)$ (probability density) vs $x$ (cell number) for different values of the multiplicative noise intensity $D$. $a=1$, $b=0.1$, $\alpha=0.5$, $\lambda=0.0$. $D=0.3$, 0.5, 0.7, and 1.0, respectively (units are arbitrary).

Figures 3 and 4 show the effect of the strength of noise $\epsilon(t)$ and $\Gamma(t)$ on the SPDF. When the degree of correlation of noises and the strength of the multiplicative noises are fixed, as the additive noise intensity $\alpha$ is increased, the maximum value on small value of $x$ increases and the maximum value on large value of $x$ decreases (see Fig. 3). The peak gets flattened and almost vanishes for a large enough value of $\alpha$, indicating that the additive noise is a diffusive term. The position of the extrema of the SPDF is weakly affected by the strength $\alpha$ of the additive noise. A different curve was represented when $\lambda$ and $\alpha$ are fixed and we change the multiplicative noise intensity $D$ (see Fig. 4). As $D$ is increased, the maximum of SPD moves from a large value of $x$ to small values of $x$, showing that the multiplicative noise is a drift term, which denotes that the multiplicative noise can push the system cell toward extinction. In other word, intensive fluctuation of the growth rate may cause tumor extinction.

## IV. CONCLUSIONS

In summary, we have studied the effects of environmental fluctuations on tumor cell growth and its steady state properties. For large values of $\lambda$ the distribution of cell population is peaked at $x=0$, which denotes a high extinction rate. The additive noise is a diffusive factor, while the multiplicative noise gives a drift factor in the process. It is found that environmental intensive fluctuations may cause tumor cell extinction.